\documentclass[aps,prl,twocolumn,superscriptaddress,showpacs,floatfix]{revtex4-2}
\usepackage{graphicx,bm,epsfig,textcomp,color,dcolumn,setspace,array,calrsfs,mathrsfs,amsmath,amssymb,mathptmx,gensymb,booktabs,url,bibentry,natbib,subfigure,float,braket,lipsum,xcolor,relsize}
\usepackage[T1]{fontenc}
\usepackage[mathscr]{eucal}
\usepackage[bookmarks=true,colorlinks,linkcolor=blue,urlcolor=blue,citecolor=blue,plainpages=false,pdfpagelabels,final,breaklinks=true]{hyperref}
\usepackage{cleveref}
\DeclareMathAlphabet{\mathcal}{OMS}{cmsy}{m}{n}

\begin{document}
\title{Proper Theory of Magnon Orbital Angular Momentum at Equilibrium}
\author{Junyu Tang}
\affiliation{Department of Physics and Astronomy, University of California Riverside, California 92521, USA}
\author{Ran Cheng}
\affiliation{\protect\mbox{Department of Electrical and Computer Engineering, University of California, Riverside, California 92521, USA}}
\affiliation{Department of Physics and Astronomy, University of California Riverside, California 92521, USA}
\affiliation{Department of Materials Science and Engineering, University of California, Riverside, California 92521, USA}

\begin{abstract}
The orbital motion of chargeless bosons, unlike that of electrons, does not generate a magnetic moment and thus cannot directly interact with magnetic fields. To formulate the orbital angular momentum (OAM) of magnons, we first identify its proper conjugate variable by considering the Aharonov-Casher effect, which gives rise to a virtual perturbation to the equilibrium state, allowing us to calculate the magnon OAM as a virtual response to an infinitesimal electric field divergence. At finite temperatures, both self-rotation and topological contributions to the magnon OAM are explicitly derived, analogous to their electronic counterpart but with the correct bosonic statistics. In a two-dimensional honeycomb lattice, we show that the Dzyaloshinskii–Moriya interaction induces a large magnon OAM in both the ferromagnetic and antiferromagnetic phases. Our formalism can be generalized to other chargeless bosons with intrinsic spin.
\end{abstract}

\maketitle

\textit{Introduction.}---While the modern theory of the electron’s orbital angular momentum (OAM) had been established about two decades ago~\cite{Xiao_PRL_2005,Thonhauser_PRL_2005,Shi_PRL_2007}, it was not until recently that the orbital degree of freedom can be manipulated for real applications. Physical phenomena such as the orbital Hall effect~\cite{Choi_Nature_2023,Tanaka_PRB_2008}, orbital Nernst effect~\cite{Salemi_PRB_2022,Salemi_PRM_2022}, orbital Edelstein effect~\cite{Hamdi_NatPhys_2023,Yoda_NanoLett_2018}, and other orbital-related effects, which parallel their spin counterparts, have been fostering an emerging playground of ``orbitronics”~\cite{Bernevig_PRL_2005,Dongwook_EuroLetter_2021}. In contrast to electrons, however, magnons (quanta of magnetic precessions) are charge-neutral bosons, whose orbital motion does not generate a magnetic moment, obscuring its physical implications.

Robin \textit{et al.}~\cite{Robin_PRL_2020} interpreted the magnon’s orbital magnetic moment as the deviation of the overall thermodynamic magnetization from the magnon spin magnetization. Such difference originates from the slight canting of the magnetic ground state induced by a magnetic field, which cannot be attributed to the magnon OAM due to the charge-neutral nature of magnons. Fishman \textit{et al.}~\cite{Fishman_PRL_2022} defined the zero-temperature magnon OAM utilizing the canonical momentum that bears no gauge invariance, thus the resulting OAM does not directly correspond to a physical observable. A remedy for gauge-invariance resorts to the decomposition of crystal-momentum integral into the radial and angular parts~\cite{Fishman_PRB_2023}. Nonetheless, the angular integral is not naturally compatible with the integration over the whole Brillouin zone (BZ) whose geometry may not be rotationally symmetric in general. Moreover, owning to their bosonic statistics, magnons are thermally populated only at finite temperatures, so the magnon OAM should 
vanish at zero temperature. Another common practice of exploring the magnon OAM has been the direct application of the electron OAM formalism to magnons~\cite{Gyungchoon_NanoLetter_20242024,Jinyang_NanoLett_2025,Matsumoto_PRL_2011} regardless of their fundamental distinctions, which seems to be questionable and is short of a proper justification.

In this letter, we develop a proper theory of the magnon OAM at thermal equilibrium from a thermodynamic perspective within linear response regime, which is manifestly gauge invariant at every $\bm{k}$ point, without relying on any prescribed symmetry of the BZ. Starting with a heuristic classical picture, we first demonstrate that the orbital motion of a magnon directly couples to the divergence of an electric field via the Aharonov–Casher (AC) effect~\cite{Aharonov_PRL_1984}, which allows us to define the OAM of an individual classical magnon. Under this simple picture, we identify the in-plane divergence of an electric field---rather than the field itself---as the thermodynamic conjugate to the magnon OAM. Next, we consider a system of magnons that are subject to an electric field inhomogeneity as a \textit{virtual} perturbation (which goes to zero at the end), so the magnon OAM can be unambiguously derived as the corresponding virtual response. The results are categorized into two distinct contributions: a self-rotation term and a topological term, analogous to those for electrons. Our formalism remains valid for finite temperatures as long as the magnetic ground state is long-range ordered. As an example, we calculate the magnon OAM in a honeycomb lattice affording the Dzyaloshinskii–Moriya interaction (DMI)~\cite{Dzyaloshinsky_JPCS_1958,Moriya_PR_1960}. We find that the DMI could induce a large magnon OAM in both the ferromagnetic (FM) and antiferromagnetic (AFM) phases. Our theory timely clarify a controversial issue, laying the foundation for studying the OAM of other chargeless bosons carrying intrinsic spins.

\begin{figure}[t]
\centering
\includegraphics[width=1\linewidth]{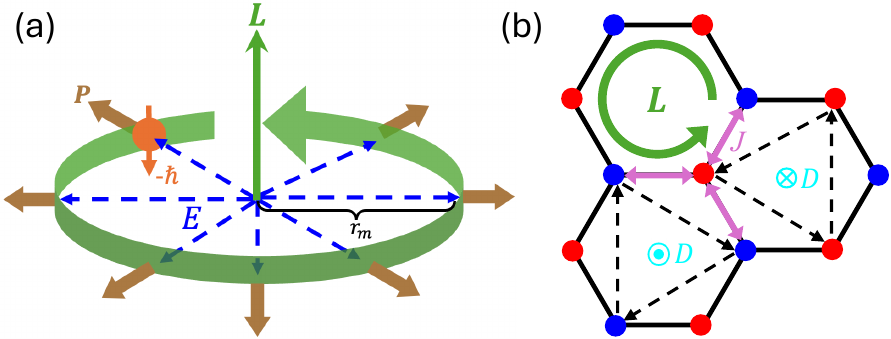}
\caption{(a) Schematics of the magnon orbital motion (green arrow), which generates a relativistic electric polarization $\bm{P}\propto \bm{v}\times\hat{\bm{z}} $ (brown arrow) that couples to the probe electric field (blue dashed arrow) with finite in-plane divergence. (b) Honeycomb lattice with $A$ and $B$ sublattices shown in red and blue, respectively. Pink arrows denote nearest-neighbor bonds with exchange interaction $J$, while black dashed arrows denote next-nearest-neighbor bonds carrying a
DMI interaction, whose sign depends on the chirality of the bond. The green loop illustrates one representative magnon orbital motion; in general, such an orbit need not be confined to a single plaquette and may extend over several unit cells.}\label{fig:model}
\end{figure}

\textit{Intuitive picture}.---In a periodic crystal, a major difficulty in defining OAM is that the naive thermal expectation of a temperature-independent OAM operator $\braket{\hat{\bm{L}}}=\mathrm{Tr}[\rho \hat{\bm r}\times\hat{\bm v}]$ is ill-defined~\cite{Thonhauser_PRL_2005,Vanderbilt_2006_PRB}. This fundamental issue carries over to magnons, motivating us to follow the spirit of the modern theory of electron OAM~\cite{Shi_PRL_2007}, where one can bypass the ill-defined position operator but instead derive the OAM thermodynamically through a virtual external perturbation. For a charge-neutral magnon whose orbital motion does not directly produce a magnetic moment like electrons, we need to first identify its correct conjugate variable. In a heuristic classical picture  [Fig.~\ref{fig:model}], a magnon orbiting on the $xy$ plane with radius $r_m$ and angular frequency $\omega$ is described by a classical vector $\bm{r}(t)=r_m[\cos(\omega t),\sin(\omega t),0]$, and carries the OAM $\bm{L}=L\hat{\bm{z}}=\bm{r}\times\dot{\bm{r}}$~\cite{mass}. Unlike electrons, the orbital motion of magnons does not couple to external magnetic fields since it does not generate a magnetic moment. Instead, we consider an external electric field $\bm{E}$, under which the above orbital motion acquires a relativistic energy correction $\Delta U$. In the co-moving frame of the magnon, $\Delta U$ represents the coupling between the magnon spin and a magnetic field $\bm{B}\approx -\dot{\bm{r}}\times \bm{E}/c^2$ with $c$ being the speed of light (note that $\bm{B}$ vanishes in the lab frame), which could be regarded as the AC effect~\cite{Aharonov_PRL_1984}. Here, we assume that the magnon velocity $\dot{\bm{r}}$ is much smaller than $c$. For a magnon of spin $-\hbar \hat{\bm{z}}$, the energy correction $\Delta U$ is
\begin{align}
    \Delta U = \frac{\gamma\hbar}{c^2} \hat{\bm{z}}\cdot (\dot{\bm{r}}\times \bm{E}) = -\bm{P}\cdot \bm{E},
    \label{eq:DeltaU}
\end{align}
where $\gamma\ (>0)$ is the gyromagnetic ratio, the electric field can be decomposed as
\begin{align}
 \bm{E}=\bm{E}^0+E_x(x,y)\hat{\bm{x}}+E_y(x,y)\hat{\bm{y}},
\end{align}
and $\bm{P}$ can be interpreted as the electric polarization of the magnon arising from its orbital motion such that
\begin{align}
    \bm{P}=\frac{\gamma \hbar}{c^2} \dot{\bm{r}}\times \hat{\bm{z}}.
    \label{eq:P_AC}
\end{align}
Note that the $E_z$ component does not concern us when the magnon moves in the $xy$ plane with its spin along $\hat{z}$. It's obvious that the uniform part $\bm{E}^0$ does not contribute to the energy correction after an average, and $E_x$ and $E_y$ should be expanded around the orbital center.

The probing electric field should have a variation scale far exceeding the lattice constant, hence it suffices to just expand $E_x$ and $E_y$ up to their first-order gradients. Eventually, we will take $r_m\rightarrow 0$ while fixing the magnon OAM amplitude $L\equiv\omega r_m^2$. We emphasis that such real-space orbital average is fundamentally different from the circular integration in the $\bm{k}$ space in Ref~\cite{Fishman_PRB_2023}. The averaged energy correction $\Delta\bar{U}$ arising from the magnon orbital motion becomes
\begin{align}\label{eq:U_ave}
    \Delta \bar{U}&=\frac{\omega}{2\pi}\int_0^{2\pi/\omega} \frac{\Delta U}{2} dt=-\frac{\gamma \hbar L }{4c^2} (\nabla_\perp\!\cdot\!\bm{E}),
\end{align}
where $\nabla_\perp\cdot \bm{E}\equiv\partial_x E_x + \partial_y E_y$. In the above equation, the prefactor $1/2$ originates from the Thomas precession, arising from the relativistic correction of the successive non-collinear boosts in a rotating (non-inertial) frame~\cite{thomas1926motion,jackson,Ouvry_2014}. Such kinematic correction, which had been applied to electron's spin-orbit coupling, is independent of particle statistics and should be applicable to the orbital motion of magnons as well. An intuitive picture revealed by Eq.~\eqref{eq:U_ave} is that the orbital motion produces an electric multipole in the lab frame that couples to $\nabla_\perp \cdot \bm{E}$ rather than $\bm{E}$ itself (Fig.~\ref{fig:model}).
Therefore, the magnon OAM can be derived from the grand canonical potential $\Omega$ by treating $\nabla_\perp\cdot \bm{E}$ as a virtual perturbation (similar to the magnetization obtained from a virtual magnetic field):
\begin{align}
    L= - \frac{4c^2}{\gamma \hbar} \lim_{\nabla_{\perp}\cdot\bm{E}\rightarrow0} \left. \frac{\partial \Omega}{\partial(\nabla_\perp \!\cdot\! \bm{E})}\right|_{T}. \label{eq:trueL}
\end{align}
Practically, it is more straightforward to calculate the internal energy $U$ than $\Omega$ from a microscopic magnon Hamiltonian, so we need to transform $L$ in Eq.~\eqref{eq:trueL} into a quantity directly related to $U=\Omega+TS+\mu N$. What serves this purpose is an auxiliary field satisfying~\cite{Shi_PRL_2007}
\begin{align}
   \tilde{\mathcal{L}} \equiv \frac{\partial(\beta L)}{\partial\beta} = -\frac{4c^2}{\gamma \hbar}\lim_{\nabla_{\perp}\cdot\bm{E}\rightarrow0}  \left. \frac{\partial U}{\partial(\nabla_\perp \!\cdot\! \bm{E})}\right|_{T} \label{eq:L_and_L},
\end{align}
where $\beta=1/k_BT$ with $k_B$ being the Boltzmann constant. Eq.~\eqref{eq:L_and_L} can be justified by making use of Maxwell's relations~\cite{Maxwell}. Since the total magnon number $N$ is not conserved, the chemical potential $\mu$ vanishes identically and $U=\Omega+TS$. From Eq.~\eqref{eq:L_and_L}, we can obtain the magnon OAM by integrating $\tilde{\mathcal{L}}$ with respect to $\beta$, i.e., $L=(1/\beta)\int\tilde{\mathcal{L}}d\beta$. In the presence of zero-point quantum fluctuations (\textit{e.g.}, the eigen-energy $\epsilon_{n\bm{k}}$ of AFM magnons), an additional temperature-independent contribution $\epsilon_{n\bm{k}}/2$ should be added to $U$. Nevertheless, the zero-point energy does not affect the magnon OAM as explained in the supplementary materials (SM)~\cite{SM}.

\textit{Quantum perturbation formalism.}---To avoid the nonlocality of OAM when applying the perturbation theory, we must consider an infinitely slow spatial variation of $\bm{E}$. Therefore, we now consider a specific field profile
\begin{align}
    \bm{E}(\bm{r})=\frac{\rho}{2q}[\sin(qx)\hat{\bm{x}}+\sin(qy)\hat{\bm{y}}],
\end{align}
controlled by a wave vector $q$ which will tend to zero at the end. Since in-plane divergence $\nabla_\perp\cdot\bm{E}=\rho[\cos(qx)+\cos(qy)]/2$ reduces to $\rho$ in the limit of $q\rightarrow 0$, we should also take $\rho\rightarrow0$ to conform with Eqs.~\eqref{eq:trueL} and~\eqref{eq:L_and_L}. By construction, the above field profile satisfies $\partial_x E_y\pm \partial_y E_x=0$, (i.e., $\nabla\times\bm{E}=0$), ensuring that no additional (time-dependent) magnetic field  $B_z$ is generated per Faraday's effect, hence the response of internal energy $\delta U$ is solely attributed to $\nabla_{\perp}\cdot\bm{E}$ at $q\rightarrow 0$. At this point, $\tilde{\mathcal{L}}$ can be readily obtained by extracting the corresponding Fourier component of $\delta U$ as
\begin{align}
    \tilde{\mathcal{L}} = -\frac{4c^2}{\gamma \hbar}\lim_{q,\rho\rightarrow 0}\frac{2}{\rho V}\int dxdy\ \delta U [\cos(qx)+\cos(qy)],
    \label{eq:Fourier}
\end{align}
where $V$ is the unit cell area.

We now quantify $U$ by the microscopic Hamiltonian $\hat{H}$ through $U=\sum_{n\bm{k}} b_{n\bm{k}}\bra{\tilde{\psi}_{n\bm{k}}} \hat{H}\ket{\tilde{\psi}_{n\bm{k}}}$, where $b_{n\bm{k}}$ is the Bose-Einstein distribution function and $\ket{\tilde{\psi}_{n\bm{k}}}$ is the perturbed magnon eigenfunction of band $n$ and momentum $\bm{k}$. Decomposing $\hat{H}=\hat{H}_0+\delta \hat{H}$ and $\ket{\tilde{\psi}_{n\bm{k}}}=\ket{\psi_{n\bm{k}}} + \ket{\delta \psi_{n\bm{k}}}$ with $\hat{H}_0$ and $\ket{\psi_{n\bm{k}}}$ being the unperturbed quantities and $\delta$ representing virtual perturbations, we obtain
\begin{align}
    \delta U = &\sum_{n\bm{k}}\delta b_{n\bm{k}}\bra{\psi_{n\bm{k}}} \hat{H}_0\ket{\psi_{n\bm{k}}} + b_{n\bm{k}}\bra{\psi_{n\bm{k}}} \delta \hat{H}\ket{\psi_{n\bm{k}}}\nonumber\\ 
    &\qquad +\left[ b_{n\bm{k}}\bra{\psi_{n\bm{k}}} \hat{H}_0\ket{\delta\psi_{n\bm{k}}}+ c.c\right]
    \label{eq:deltaU}
\end{align}
as the first-order correction of the internal energy, which contains contributions from the changes of the distribution function $\delta b_{n\bm{k}}$, the Hamiltonian $\delta \hat{H}$, and the eigenfunction $\ket{\delta \psi_{n\bm{k}}}$, respectively.

According to Eq.~\eqref{eq:DeltaU}, the perturbation Hamiltonian takes the form $\delta \hat{H}=-\hat{\bm{P}}\cdot \bm{E}(\bm{r})$ where $\hat{\bm{P}}(\bm{k})=\gamma \hbar \hat{\bm{v}}(\bm{k})\times \hat{\bm{z}}/c^2 $ with $\hat{\bm{v}}=\partial \hat{H}_0(\bm{k})/\partial(\hbar\bm{k})$ being the velocity operator. By defining $\ket{u_{n,\bm{k}}}=e^{-i\bm{k}\cdot\bm{r}}\ket{ \psi_{n\bm{k}}}$, we can express the perturbed wavefunction as
\begin{widetext}
\begin{align}
    \ket{\delta \psi_{n\bm{k}}}=\frac{-\rho}{8iq}\sum_{m}\left[
    \frac{e^{i(\bm{k}+q\hat{\bm{x}})\cdot\bm{r}} \bra{u_{m,\bm{k}+q\hat{\bm{x}}}} \hat{P}_x(\bm{k}+q\hat{\bm{x}}) + \hat{P}_x(\bm{k})  \ket{u_{n,\bm{k}}}}
    {\epsilon_{n,\bm{k}}-\epsilon_{m,\bm{k}+q\hat{\bm{x}}}}\ket{u_{m,\bm{k}+q\hat{\bm{x}}}}-(q\rightarrow -q) \right] + (\hat{\bm{x}}\rightarrow \hat{\bm{y}}),\label{eq:delta_psi}
\end{align}
\end{widetext}
where $\epsilon_{n\bm{k}}$ is the unperturbed magnon energy. To simplify the notation, we will omit the $\bm{k}$ index in $\epsilon_{n\bm{k}}$ and $b_{n\bm{k}}$ in the following.

As detailed in the SM~\cite{SM}, the first two terms of Eq.~\eqref{eq:deltaU} turn out to be zero, while only the last term (arising from $\delta\psi_{n\bm{k}}$) survives. Inserting Eq.~\eqref{eq:delta_psi} into Eq.~\eqref{eq:deltaU} and then into Eq.~\eqref{eq:Fourier}, we obtain $\tilde{\mathcal{L}} = \tilde{\mathcal{L}}_1 + \tilde{\mathcal{L}}_2$ with
\begin{subequations}\label{eq:L1L2}
\begin{align}
    \tilde{\mathcal{L}}_1 &= \frac{2c^2}{\gamma} \sum_{\bm{k},m\neq n} \epsilon_{m}(b_{m}-b_{n}){\rm Im}\left[\frac{ \bm{v}_{nm}\cdot \bm{P}_{mn}}{(\epsilon_{n}-\epsilon_{m})^2}\right],\\
    \tilde{\mathcal{L}}_2 &= \frac{2c^2}{\gamma} \sum_{\bm{k},m\neq n} \epsilon_{n} \frac{\partial b_{n}}{\partial \epsilon_n} {\rm Im}\left[\frac{ \bm{v}_{nm}\cdot \bm{P}_{mn}}{(\epsilon_{n}-\epsilon_{m})}\right],
\end{align}
\end{subequations}
where the matrix elements read $\bm{v}_{nm}=\bra{u_{n,\bm{k}}} \hat{\bm{v}}(\bm{k})\ket{u_{m,\bm{k}}}$ and $\bm{P}_{nm}=\bra{u_{n,\bm{k}}} \hat{\bm{P}}(\bm{k})\ket{u_{m,\bm{k}}}$. For electrons, the energy factor $\epsilon_n$ in $\tilde{\mathcal{L}}_2$ is $(\epsilon_{n} -\mu)$, resulting in a vanishing $\tilde{\mathcal{L}}_2$ at zero temperature as enforced by the Fermi-Dirac distribution function $f_{n}$ satisfying $\frac{\partial f_{n}}{\partial \epsilon}\approx\delta(\epsilon_{n} -\mu)$ for $T\to0$. However, this is not true for magnons (in fact, bosons in general); and $\tilde{\mathcal{L}}_2$ could be even larger than $\tilde{\mathcal{L}}_1$. 

Finally, by integrating $\tilde{\mathcal{L}}$ over $\beta$ based on Eq.~\eqref{eq:L_and_L}, we obtain the magnon OAM at finite temperature as
\begin{align}
&L = -\sum_{\bm{k},n}  [\ell_n(\bm{k}) b_n-\frac{4}{\beta \hbar }\Omega_n(\bm{k}) \ln (1-e^{-\beta\epsilon_{n}}) ],
\label{eq:L}
\end{align}
where we have divided the results into two suggestive parts: $\ell_n(\bm{k})$ can be intuitively interpreted as the self-rotation of a magnon wavepacket; and a topological correction related to the Berry curvature $\Omega_n(\bm{k})$. In two dimensions, they are
\begin{subequations}\label{eq:L_and_B}
\begin{align}
&\ell_n(\bm{k}) = -\frac{2}{\hbar}{\rm Im} \left[\bra{\partial_{\bm{k}} u_{n,\bm{k}}}(\epsilon_n-\hat{H}_0)\times \ket{\partial_{\bm{k}} u_{n,\bm{k}}}\right]_z,\\
&\Omega_n(\bm{k}) = -{\rm Im} \left[\bra{\partial_{\bm{k}} u_{n,\bm{k}}}\times \ket{\partial_{\bm{k}} u_{n,\bm{k}}}\right]_z,
\label{eq:Berry}
\end{align}
\end{subequations}
which are manifestly gauge invariant. Equations~\eqref{eq:L} and~\eqref{eq:L_and_B} are the central results of this work~\cite{factor}, which are applicable to not just magnons but any chargeless bosons with intrinsic spin. As an independent check, we provide in the SM~\cite{SM} an alternative derivation of the same formulae using the semiclassical wave packet formalism.

Interestingly, the topological term in Eq.~\eqref{eq:L} vanishes at $T\rightarrow0$, contrasting sharply to the OAM of electrons whose topological term reduces to $\Omega_n f_n$ as $T\rightarrow0$. We note that the topological term does not have a similar simple form $\Omega_n b_n$ at zero-temperature limit~\cite{Matsumoto_PRL_2011}. The similarity of the self-rotation term between electrons and magnons can be attributed to the hybrid product structure introduced by the $\bm{P}$ operator and $\nabla_{\perp}\cdot\bm{E}$~\cite{hybrid}.

To further check the validity of our results, we now take the spatial curl $\nabla\times$ of the magnon OAM, intentionally creating a non-equilibrium system driven by an inhomogeneous stimulus. Akin to the electronic case~\cite{Xiao_PRL_2006}, here $\bm{\nabla}\times L$ bears resemblance to a magnetization current (while $L$ is not accompanied by a magnetic moment) that relates the equilibrium OAM to a non-equilibrium transverse current $\bm{J}_{\rm trans}$, as if the system is perturbed by a temperature gradient $\bm{\nabla}T$. Disregarding the front factor, we have
\begin{align}\label{eq:MSN}
    \bm{J}_{\rm trans}\propto \bm{\nabla}\times L\propto \sum_{n,\bm{k}}\Omega_{n}(\bm{k}) [\hat{z}\times\bm{\nabla}T] c_1(b_{n}), 
\end{align}
where $c_1(x)\equiv x\ln(x)-(1+x)\ln(1+x)$, $\bm{J}_{\rm trans}$ turns out to be nothing but the magnon spin Nernst current~\cite{Cheng_PRL_2016,Vladimir_PRL_2016}. Compared with previous studies of the magnon OAM~\cite{Robin_PRL_2020,Fishman_PRL_2022,Matsumoto_PRL_2011}, only our theory can reproduce the correct expression of the magnon spin Nernst effect.

\textit{Examples.}---Consider a honeycomb lattice with nearest-neighbor Heisenberg exchange interaction $J$ as illustrated in Fig.~\ref{fig:model}. For the next-nearest neighbors, the broken inversion symmetry allows for the DMI~\cite{Cheng_PRL_2016}. The minimal magnetic Hamiltonian is~\cite{Hamiltonian}
\begin{align}
    H=&J\sum_{\langle i j\rangle} \bm{S}_i\cdot \bm{S}_j + \sum_{\langle\langle i,j\rangle\rangle} \bm{D}_{ij}\cdot (\bm{S}_i\times \bm{S}_j) \notag\\
    &\qquad+ \sum_{i} \left[\gamma  B_z S^z_i -  A_z(S^z_i)^2\right],
\end{align}
where the DMI vector $\bm{D}_{ij}=\pm D\hat{z}$ points along $\hat{\bm{z}}$ and its sign is determined by the chirality of the $i-j$ bond. The on-site term contains the Zeeman energy from an external magnetic field $B_z$ and the easy-axis anisotropy ($A_z>0$). Under the Holstein–Primakoff transformation, the Hamiltonian can be written as $H=\frac{1}{2}\sum_{\bm{k}}\Psi^\dagger _{\bm{k}}H_{\bm{k}}\Psi_{\bm{k}}$, where $H_{\bm{k}}$ is
\begin{align}
    H_{\bm{k}}=
    \begin{bmatrix}
    \mathcal{D}_0+\mathcal{D}_1  & \Delta^\dagger & 0 & T^\dagger_+ \\
    \Delta & \mathcal{D}_0+\mathcal{D}_2 & T_-  & 0\\
    0 & T^\dagger_-  & \mathcal{D}_0+\mathcal{D}_3  &  \Delta^\dagger \\ 
    T_+   & 0 & \Delta & \mathcal{D}_0+\mathcal{D}_4
\end{bmatrix}, \label{eq:Hk}
\end{align}
and the Nambu basis is $\Psi = (a_{\bm{k}}, b_{\bm{k}}, a_{-\bm{k}}^\dagger, b_{-\bm{k}}^\dagger )^T$ with $a_{\bm{k}}$ ($a_{\bm{k}}^\dagger$) and $b_{\bm{k}}$ ($b_{\bm{k}}^\dagger$) the magnon annihilation (creation) operators on the $A$ and $B$ sites. The specific expressions of each parameter in the $H_{\bm{k}}$ for the FM ($J<0$) and AFM ($J>0$) phases are laid out in the SM~\cite{SM}. To properly apply Eq.~\eqref{eq:Berry} while preserving the bosonic commutation relations, we employ Colpa’s method~\cite{Colpa_Physica_1978} for para-unitary diagonalization of $H_{\bm{k}}$.

An electric field could induced spin-dependent phase in the electron's hopping via the AC effect, manifesting as an effective DMI~\cite{Liu_2011_PRL,Nagaosa_2005_PRL}. From the modified energy, one can read off an effective polarization operator $\hat{\bm{P}}_{\rm eff}\propto\sum_{i,j}J_{i,j}\bm{a}_{ij}\times({\bm{S}}_i\times {\bm{S}}_j)$ with $J_{i,j}$ and $\bm{a}_{ij}$ the corresponding exchange interaction and the bond vector connecting the $i$-th and $j$-th sites. It can be shown that $\hat{\bm{P}}_{\rm eff}$ is equivalent to $\bm{P}$ given by Eq.~\eqref{eq:P_AC} when the intrinsic DMI vanishes~\cite{p_operator}.

\begin{figure}[t]
\centering
\includegraphics[width=1.0\linewidth]{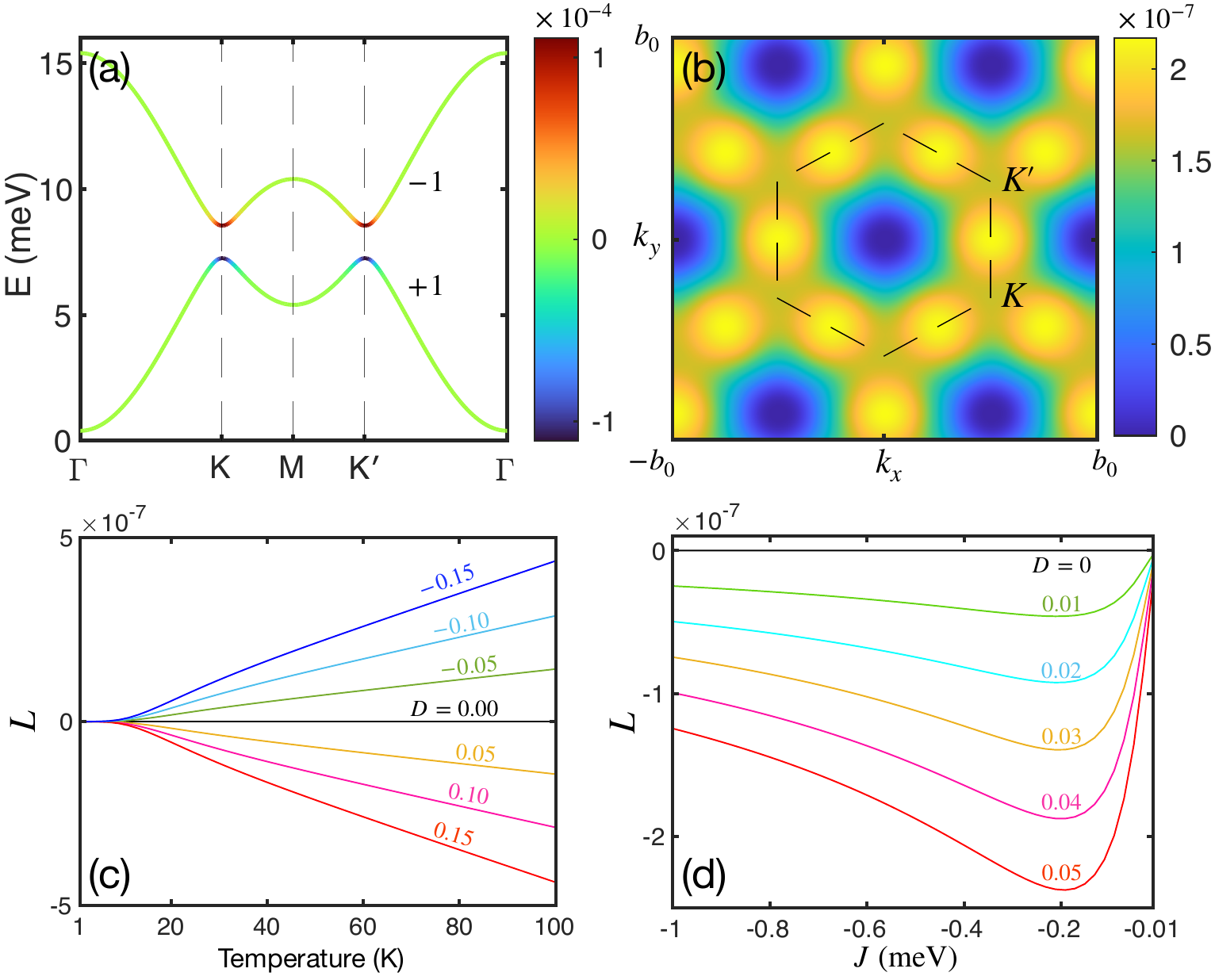}
\caption{(a) FM magnon band structure (weighted by magnon OAM). Integers represent the Chern numbers of the corresponding bands. (b) the $k$-space distribution of total FM magnon OAM. The dashed lines mark the first BZ with $b_0$ the length of reciprocal lattice. (c) FM Magnon OAM $L$ per unit cell as a function a temperature and (d) exchange interaction for different DMI. In all figures, magnon OAM are plotted in the unit of $\rm m^2/s$.}\label{fig:FM_OAM}
\end{figure}

\textit{FM phase.}---Using a set of parameters $S=5/2,\ T=100\ {\rm K},\ B_z=0$ and $\ J=-1,\ A_z=0.1, \ D= 0.05$ (all in meV unit), we plot the band structure weighted by the magnon OAM in Fig.~\ref{fig:FM_OAM}(a). Notably, FM magnons exhibit a large OAM (on the order of $10^{-4}\rm \ m^2/s$) around $K$ and $K'$ points where gaps are open by a finite $D$. However, the two bands carry opposite OAM, so the total magnon OAM is reduced to the order of $10^{-7}\rm \ m^2/s$, maximizing around the $M$ points as shown in Fig.~\ref{fig:FM_OAM}(b). The temperature dependence of the magnon OAM is plotted in Fig.~\ref{fig:FM_OAM}(c), where we observe that a noticeable OAM starts to develop around 10 K and increases monotonically with the temperature, consistent with the bosonic nature of magnons. Here, we assume that 100 K is below the Curie temperature so that the system retains a well-defined FM order in the ground state. Additionally, we find that the magnon OAM is an odd function of the DMI and increases with a growing magnitude of $D$. Moreover, as plotted in Fig.\ref{fig:FM_OAM}(d), the magnon OAM exhibits a non-monotonic dependence on the exchange interaction $J$, reaching its maximum around $J=-0.2\ \rm meV$. This behavior can be attributed to the competition between the magnon energy and velocity: as $J$ increases, the magnon velocity rises up while the magnon energy also increases, leading to a reduction of population~\cite{FM_stable}. Since both bands are subject to the same Zeeman energy shift, a finite $B_z$ can only lead to a slight change of the magnon OAM through the distribution function.

\begin{figure}[t]
\centering
\includegraphics[width=1.0\linewidth]{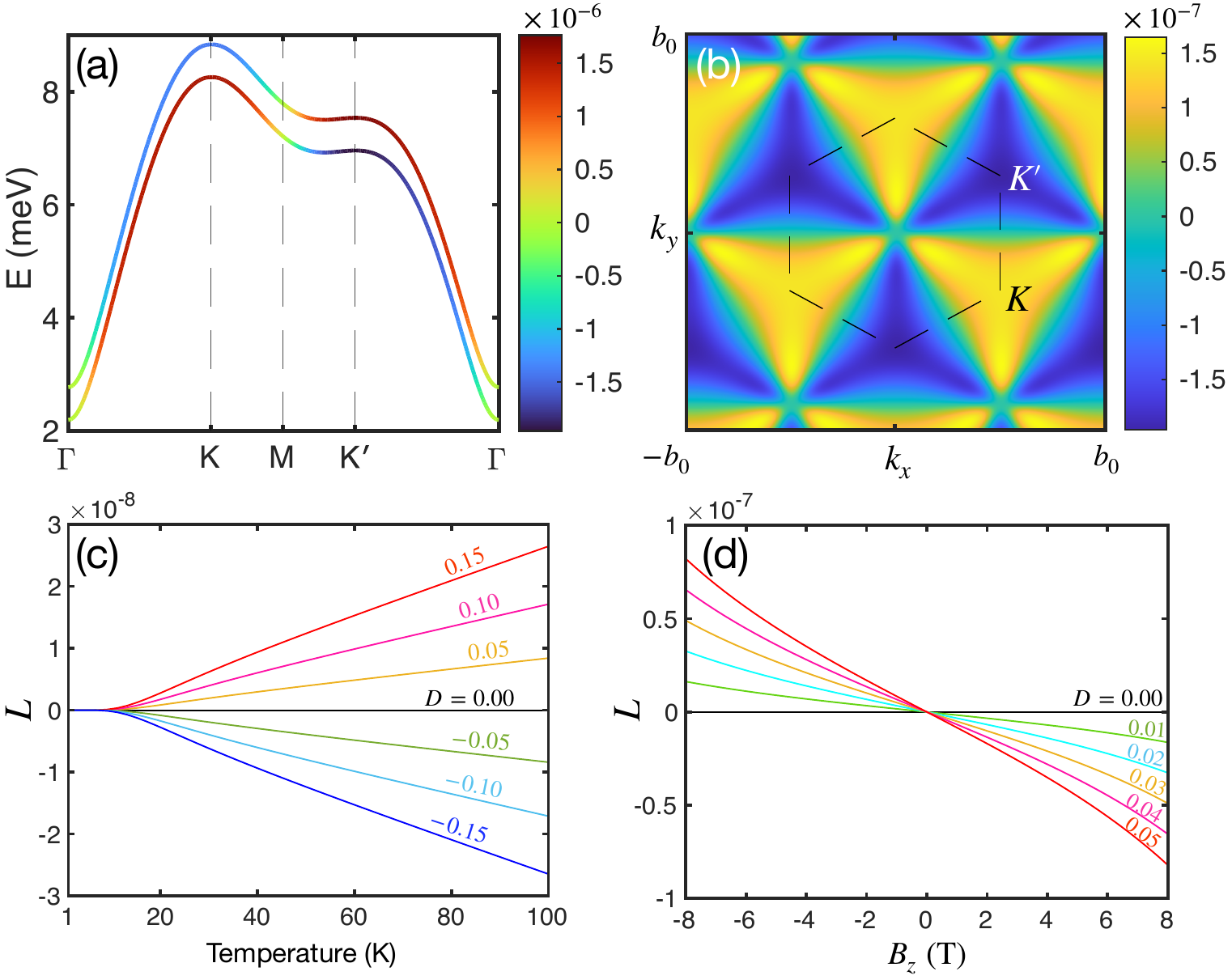}
\caption{(a) AFM magnon band structure (weighted by magnon OAM). Both bands have zero Chern number. (b) the $k$-space distribution of total AFM magnon OAM. (c) AFM Magnon OAM $L$ per unit cell as a function a temperature and (d) magnetic field for different DMI. In all figures, magnon OAM are plotted in the unit of $\rm m^2/s$. }\label{fig:AFM_OAM}
\end{figure}

\textit{AFM phase.}---In this case, we change the exchange interaction and the magnetic field into  $J=+1\ \rm meV$ and $B_z=-1 \rm\ T$. Figure~\ref{fig:AFM_OAM}(a) shows the band structure weighted by the magnon OAM, where the AFM magnon OAM remains finite even far away from the $K$ and $K'$ points, contrasting their FM counterparts. When contributions from both bands are taken into account, the AFM magnon OAM are maximized around the $K$ and $K'$ points as plotted in Fig.~\ref{fig:AFM_OAM}(b), which shares a similar pattern with the Berry curvature [see Fig.S1(b)]. After an integration over the first BZ, the overall AFM magnon OAM is reduced to the order of $10^{-8}\ \rm m^2/s$. As shown in Fig.~\ref{fig:AFM_OAM}(c), the temperature dependence is quite similar to the FM case. It should be noted that the Berry curvature of AFM magnons in the honeycomb lattice considered here is independent of $D$~\cite{zhang2022perspective}, whereas a nonzero magnon OAM does require a finite DMI. Furthermore, since DMI alone cannot break spin degeneracy in the AFM case, a finite $B_z$ is always needed to separate the two bands of opposite spins in order to acquire a net OAM. As the applied magnetic field increases (while remaining below the spin-flop threshold), the energy gap between the two bands grows, leading to an increasing magnon OAM. This trend holds for both negative and positive $B_z$ as shown in Fig.~\ref{fig:AFM_OAM}(d).

\textit{Numerical estimation.}---We propose a polarization measurement to detect the magnon OAM. While the polarization induced by the magnon orbital motion cancels out in the bulk, a net polarization density $\bm{p}$ should be detectable on the edges~\cite{Gyungchoon_NanoLetter_20242024}. Note that Eq.~\eqref{eq:P_AC} corresponds to the total polarization $\bm{P}=\bm{p} A_m d_m$, where $A_m$ is the unit cell area. Although we are studying a 2D honeycomb lattice, it's reasonable to consider the thickness $d_m\approx 0.5\ \rm nm$ to be the typical interlayer distance of Van der Waals materials. For FM magnon at room temperature with $D=0.05$ meV, the OAM amplitude is around $L\approx 4\times10^{-7}\ \rm m^2/s$, which results in a polarization density $\bm{p}$ on the order of $10^{-4}\ {\rm C/m^2}$, provided a typical spin-orbit coupling of eV scale~\cite{Liu_2011_PRL,Gyungchoon_NanoLetter_20242024,Jing_PRL_2023}. As for the AFM magnons, a finite magnetic field is required to break the degeneracy of the two bands for a nonzero OAM. At room temperature, with $D=0.05$ meV and $B_z=-1\ \rm T$, the polarization density $\bm{p}$ is on the order of $10^{-5}\ {\rm C/m^2}$, which gives an effective magnetoelectric polarizability $\alpha_{\rm eff}\equiv |\bm{p}/B_z|$ on the order of $10^{-5}\ {\rm C/m^2\cdot T}$. For comparison, the quantized topological magnetoelectric polarizability $\alpha_{\theta}=\theta e^2/2\pi h$ of an axion insulator ($\theta=\pi$) is of the same order as $\alpha_{\rm eff}$, highlighting the significance of the magnon-OAM-induced magnetoelectric effect.

In summary, we have demonstrated that magnon OAM couples to an inhomogeneous electric field via the AC effect and formulated a true gauge-invariant theory to properly quantify the magnon OAM at finite temperatures. In a honeycomb lattice with isotropic exchange interaction, we find that a nonvanishing magnon OAM requires a finite DMI and increases monotonically with the temperature. 
Looking ahead, it would be interesting to explore magnon OAM in altermagnets, where spin splitting occurs intrinsically without an external magnetic field, and to investigate how magnon OAM responds to other stimuli with our proper formalism, such as a temperature gradient (i.e., the magnon orbital Nernst effect).

\begin{acknowledgments}
This work is supported by the National Science Foundation under Award No. DMR-2339315. We acknowledged fruitful discussions with Junren Shi and Hantao Zhang. 
\end{acknowledgments}

\bibliography{ref.bib}

\end{document}